\documentclass[twocolumn,amsmath,amssymb,prd,floatfix,nofootinbib,superscriptaddress]{revtex4}
\usepackage{graphicx}
\usepackage{bm}
\usepackage{amsmath}
\usepackage[dvipdfmx]{color}
\input{colordvi.tex}
\usepackage{color}
\usepackage{hyperref}
\usepackage{natbib}
\usepackage{url}

\newcommand{\simgt}{\lower.5ex\hbox{$\; \buildrel > \over \sim \;$}}
\newcommand{\simlt}{\lower.5ex\hbox{$\; \buildrel < \over \sim \;$}}

\begin{document}

\title[Supersonic motion]{Effect of supersonic relative motion between baryons\\ and dark matter on collapsed objects}

\author{Shinsuke Asaba}\email{asaba.shinsuke@j.mbox.nagoya-u.ac.jp}
\affiliation{Department of Physics, Graduate School of Science, Nagoya
University, Aichi 464-8602, Japan
}

\author{Kiyotomo Ichiki}
\affiliation{Department of Physics, Graduate School of Science, Nagoya
University, Aichi 464-8602, Japan
}
\affiliation{Kobayashi-Maskawa Institute for the Origin of Particles and the Universe, Nagoya University,
Aichi 464-8602, Japan}

\author{Hiroyuki Tashiro}
\affiliation{Department of Physics, Graduate School of Science, Nagoya
University, Aichi 464-8602, Japan
}

\date{\today}

\begin{abstract}
Great attention is given to the first star formation and the epoch of reionization as main targets of planned large radio interferometries~(e.g. Square Kilometre Array).
Recently, it is claimed that the supersonic relative velocity between baryons and cold dark matter can suppress the abundance of first stars and impact the cosmological reionization process. 
Therefore, in order to compare observed results with theoretical predictions
it is important to examine the effect of the supersonic relative motion on the small-scale structure formation.
In this paper, we investigate this effect on the nonlinear structure
 formation in the context of the spherical collapse model
in order to understand the fundamental
physics in a simple configuration.
We show the evolution of the dark matter sphere with
 the relative velocity by both using N-body simulations and numerically calculating the equation of motion for the dark matter mass shell.
The effects of the relative motion in the spherical collapse model appear as the delay of the collapse
 time of dark matter halos and the decrease of the baryon mass fraction
 within the dark matter sphere.
Based on these results,
we provide the fitting formula of the critical density contrast for collapses with the relative motion effect
and calculate the mass function of dark matter halos in the Press-Schechter formalism.
As a result, the relative velocity decreases the abundance of dark matter halos whose mass is smaller than
$10^8~M_\odot/h$.
\end{abstract}

\maketitle

\section{INTRODUCTION}\label{A}
The standard cosmological model, called the $\Lambda$CDM model, composed with
two relativistic species~(photons and neutrinos), two nonrelativistic matters~(baryons and dark matter), and the energy having negative pressure~(dark
energy) with a nearly scale-invariant spectrum of curvature
perturbations, has achieved great success in explaining large-scale
cosmological observations, e.g.,~large-scale structure formation~\cite{2014arXiv1411.1074A} and cosmic microwave background~\cite{2015arXiv150201582P}.

The theoretical description of small-scale structure formation is, however, still debatable.
Understanding of small-scale structure formation at high redshifts 
is essential to study first stars and the epoch of reionization~(EoR).
One expects redshifted 21~cm lines from the hyperfine structure of
hydrogen atoms as the powerful probe for the EoR and first stars~\cite{2006PhR...433..181F,2012RPPh...75h6901P} 
and the matter density underlying HI distribution constrains extended parameters of the $\Lambda$CDM model~\cite{2013PhLB..718.1186O,2014JCAP...09..014K,2014PhRvD..90h3003S}.
Currently, to probe such small-scale structure formation, there are many planed
 observations including Murchison Widefield Array~\cite{MWA} and Square Kilometre Array~(SKA)~\cite{SKA}.
Therefore, nowadays, the detailed studies on small-scale structure formation at
high redshifts attract a lot of attention.
 
Recently, Ref.~\cite{2010PhRvD..82h3520T} reports the importance
of the supersonic relative motion between dark matter and baryons on
small-scale structure formation related to the EoR.
This supersonic relative motion is originated from the difference in the
motions between baryons and dark matter before recombination.
Baryons before recombination are tightly coupled with photons by Thomson scattering.
As a result, baryons and photons act as one fluid with the sound speed $\sim c/\sqrt3$ and
have the velocity field associated with the acoustic oscillation.
On the other hand, dark matter does not suffer from Thomson
scattering and dark matter density fluctuations can grow gravitationally.
Therefore, the relative motion between baryons and dark matter is induced.
After recombination, baryons are fully decoupled with photons and the sound speed of baryons quickly drops to $\sim
6~{\rm km/s}$. Since the root mean square of the
relative velocity reaches~$\sim {30~\rm km/s}$ at that time, the relative velocity is about
five times larger than the sound speed of baryons.
Because the relative motion is highly supersonic, the effect on the
structure formation could be significant. 
In particular, the abundance of small dark matter halos~($M\simlt 10^7~M_\odot)$
is highly suppressed due to the supersonic relative motion.
This effect has been intensively studied by many authors with 
N-body/smoothed particle hydrodynamics~(SPH) simulations~\cite{2011ApJ...730L...1S,2012ApJ...747..128N,2013ApJ...763...27N}.
Therefore, according to the effect on the structure formation, the scenario of the cosmic reionization and the prediction of
the 21~cm line signals from the EoR could be modified from those
predicted in the 
conventional cosmological model.
The recent relevant studies are reviewed in  
Ref.~\cite{2014IJMPD..2330017F}.

So far the effects of the relative motion on the structure formation
have been studied in numerical simulations mainly, because these effects
are complicated.
However, a study with the analytical model is useful to obtain some
insights into physics involved in complicated phenomena.
Additionally the analysis of observation data with numerical simulations
generally takes enormous time and, sometimes, it seems unrealistic.
Therefore, modeling in a form which is easy to handle in analytic
studies is highly required.

In the study on the structure formation,
the halo mass function is one of the
interesting quantities. In particular,
the Press-Schechter formalism with the sphere collapse model
provides the mass function in the analytical form
which relatively agrees well with the results of N-body simulations.

In this paper, we revisit the effect of the supersonic relative motion
on small-scale structure formation in the context of the spherical collapse model by both N-body simulations and
a semianalytical way.
The effect of the relative motion on the
spherical collapse can be represented as the modification of the
critical density contrast.
We propose a fitting formula of the critical density contrast, 
as a function of the amplitude of the relative motion, the halo mass and the initial density fluctuation within dark matter halos.
We also apply this fitting formula to evaluate the mass function of
small dark matter
halos based on the Press-Schechter formalism.

This paper is organized as follows.
In Sec.~\ref{B} we review the effect of supersonic relative motion on the perturbation theory by taking into account the background velocity of baryons and construct the spherical collapse model with two components.
In Sec.~\ref{C} we describe the setup of our N-body simulation, 
and we show the results of the N-body simulations and check the reproducibility of the spherical collapse in Sec.~\ref{Ca}.
Moreover we present the change of the collapse time by supersonic
relative motion and the validity of the semianalytical model introduced
in Sec.~\ref{B}.
Section~\ref{D} is devoted to the discussion of the relative motion effect
on dark matter halos. We discuss the modification of the baryon
fraction in a dark matter halo by the relative motion, and,
providing a fitting formula of the modified collapse time~(i.e.~the critical density
contrast for the collapse).
We show the suppression of the dark matter halo abundance around
the EoR as an application of our results. Finally we summarize this paper in Sec.~\ref{E}.

\section{ANALYTICAL FORMALISM}\label{B}
In this section, we show the effect of the supersonic relative motion
between baryons and dark matter on the linear perturbation theory, and
evaluate this effect on the nonlinear growth by adopting the spherical collapse model.

\subsection{Perturbation theory}\label{Ba}
Since the supersonic relative motion between baryons and dark matter has been
analytically studied in the moving-background perturbation theory
(MBPT)~\cite{2010PhRvD..82h3520T},
we first make a brief review of the MBPT.
The MBPT introduces the background peculiar velocity which
corresponds to the relative velocity between baryons and dark matter, i.e. $\vec{v}^{~\rm bg}=\vec{v}_{bc}$.
According to the energy momentum
conservation equation with homogeneous background densities of baryons
and dark matter,
the evolution of $\vec{v}_{bc}$ is in reverse proportion to the scale factor due to the cosmic expansion. 

In the MBPT, the first order energy momentum conservation equations after
recombination are given by
\begin{align}
	\frac{d\delta_c}{dt}=&-\theta_c,\notag\\
	\frac{d\theta_c}{dt}=&-\frac{3H^2}{2}(\Omega_c\delta_c+\Omega_b\delta_b)-2H\theta_c,\notag\\
	\frac{d\delta_b}{dt}=&-\frac{i}{a}\vec{v}_{bc}\cdot\vec{k}\delta_b-\theta_b,\notag\\	
	\frac{d\theta_b}{dt}=&-\frac{i}{a}\vec{v}_{bc}\cdot\vec{k}\theta_b-\frac{3H^2}{2}(\Omega_c\delta_c+\Omega_b\delta_b)-2H\theta_b+\frac{c_s^2k^2}{a^2}\delta_b,\label{eqpt}
\end{align}
where $c_s$ is the sound velocity of the baryon fluid, the subscripts $c$ and $b$ denote cold dark matter
and baryons respectively and $\theta=ia^{-1}\nabla\cdot\vec{v}$ represents the
divergence of the peculiar velocity.
In Eq.~(\ref{eqpt}) we take the frame where the background velocity of cold dark matter is absent.
In other words, $\vec{v}^{\rm bg}_{b}=\vec{v}_{bc}$ and $\vec{v}^{\rm bg}_{c}=0$.
For simplicity, we ignore perturbations of the sound velocity
although they might affect the growth of the density fluctuation on small scales~\cite{2005MNRAS.362.1047N,2011MNRAS.418..906T,2011MNRAS.416..232N}.
Equation~(\ref{eqpt}) can be rewritten to the second order differential
equations of the density fluctuations as
\begin{align}
	\frac{d^2\delta_c}{dt^2}=&-2H\frac{d\delta_c}{dt}+\frac{3H^2}{2}(\Omega_c\delta_c+\Omega_b\delta_b),\notag\\
	\frac{d^2\delta_b}{dt^2}=&-\left(2H+2i\mu v_{bc}\frac{k}{a}\right)\frac{d\delta_b}{dt}+\frac{3H^2}{2}(\Omega_c\delta_c+\Omega_b\delta_b)\notag\\
		&-\left(c_s^2+\mu^2v_{bc}^2\right)\frac{k^2}{a^2}\delta_b,\label{eqdl}
\end{align}
where $\mu=\vec{v}_{bc}\cdot \vec{k}/|\vec{v}_{bc}||\vec{k}|$.

Equation~(\ref{eqdl}) tells us that the relative motion prevents the growth of
density fluctuations on small scales in the same way of the fluid pressure 
in the discussion of the Jeans instability.
On large scales where the relative motion does not have the preferred direction,
while the odd term of $\mu$ in the last equation of Eq.~(\ref{eqdl}) vanishes by averaging over all random directions
of the relative motion, the third term of the right-hand side is enhanced due to the existence of
the term with $v_{bc}^2$.
As a result, the effective Jeans
scale~(the suppression scale) of
Eq.~(\ref{eqdl}) becomes large due to the existence of the relative velocity.
Since the relative velocity after recombination is roughly $\langle v_{bc}^2\rangle^{1/2}\sim
5c_s$, the suppression scale for the relative motion is $k_{bc}=aH/\langle v_{bc}^2\rangle^{1/2}\sim 40~h{\rm Mpc}^{-1}$.
The corresponding mass scale for the suppression is $M_{bc}\sim 10^7~M_\odot/h$.

However, when we consider a local sufficiently small patch,
the odd term of $\mu$ cannot vanish in the patch. Instead, 
the relative motion in this patch can be assumed to be a homogeneous flow with one direction.
In this case, when the relative velocity
is larger than the Hubble flow, $\mu v_{bc}>k/aH$,
the density fluctuations inside the patch start to grow exponentially due
to the relative motion flow as shown in Eq.~(\ref{eqdl}).
The perturbation theory is not valid in this case,
and we need to
consider the effect of the relative velocity on the nonlinear growth,
e.g.,~in the spherical collapse model.

\subsection{Spherical collapse model}\label{Bb}
The spherical collapse model is a simple analytical model to investigate the
nonlinear evolution of an overdensity region.
In this model, the evolution of the overdensity region is described as
the motion of the constant
density spheres.
Let us consider the collapse of a mass shell inside which the mass is 
$M=4\pi x_i^3\bar{\rho}_i(1+\delta_i)/3$, 
where $x_i$ is the initial radius and $\delta_i$ is the initial density
contrast within the sphere with the radius $x_i$~(hereafter the subscript $i$ represents the initial time value).
The equation of motion~(EoM)~for the proper radius $x$ of a shell
is written as
\begin{align}
	\frac{d^2 x}{dt^2}=-\frac{GM}{x^2}. \label{eq1}
\end{align}
Equation~(\ref{eq1}) can be solved analytically, and the solution is given by
\begin{align}
	\tilde{t}=\frac{t}{t_i}&=\frac{3}{4\sqrt{1+\delta_i}}\left[1-\frac{(v_i/H_ix_i)^2}{1+\delta_i}\right]^{-3/2}(\theta-\sin\theta),\notag\\
	 \tilde{x}=\frac{x}{x_i}&=\frac{1}{2}\left[1-\frac{(v_i/H_ix_i)^2}{1+\delta_i}\right]^{-1}(1-\cos\theta), \label{eq3}
\end{align}
where $t_i$ is the initial time and $v_i$ is the initial velocity.	 
The solution of Eq.~(\ref{eq3}) depends only on $\delta_i$ and $v_i$.
In order to keep the constant mass $dM/dt=0$, we give the initial
velocity of the shells
\begin{align}
	v_i=H_ix_i\left[1-\frac{\delta_i}{3(1+\delta_i)}\right], \label{eq4}
\end{align}
where we assume the matter dominated era, $t\propto a^{3/2}$.
The first term represents the Hubble flow and the second term corresponds to the peculiar velocity.
According to Eq.~(\ref{eq4}), the solution, Eq.~(\ref{eq3}) depends on only the initial
density fluctuation,~$\delta_i$.
Furthermore, we can obtain the critical density contrast that is the density contrast at the collapse time $\theta=2\pi$
in the linear perturbation theory,
\begin{align}
	\delta_{\rm crit}=\frac{a(\theta=2\pi)}{a_i}\delta_i=\frac{3}{5}\left(\frac{3}{2}\pi\right)^{2/3}, \label{eq5}
\end{align}
where we use the fact that the growth factor of the matter density perturbation is proportional to the scale factor in the matter dominated era.

Next we consider the effect of the homogeneous supersonic relative motion on the
nonlinear evolution in the spherical collapse model.
A simple extension is to introduce the two kinds of mass shells for dark
matter and baryons. Taking into account the supersonic relative
motion, the baryon mass shells have the initial bulk
velocity, because we take the frame where baryons have the homogeneous
relative flow to dark matter. As a result, the collapsing of the baryon mass shells
is not spherical and these mass shells 
are collapsing to the different
position from the dark matter shells.
However, the collapse of dark matter precedes the one of the baryons in a halo formation and 
we are interested in a baryon fraction within a collapsed dark matter halo. 
Therefore, we focus on the dark matter mass shell
and, instead of following the evolution of the baryon mass shells, we
introduce the baryon mass within the dark matter mass shell,
$M_b$, and rewrite Eq.~(\ref{eq1}) to
\begin{align}
	\frac{d^2 x_c}{dt^2}&=-\frac{G(M_{c,i}+M_b)}{x_c^2},\notag\\
	M_{c,i}&=\frac{4\pi}{3} \bar{\rho}_{c,i}x_{c,i}^3(1+\delta_{c,i}),\notag\\
	M_b&=\frac{4\pi}{3}\bar{\rho}_bx_c^3(1+\delta_b),\label{rveq}
\end{align}
where $M_{c,i}$
is the mass within the shell with the initial radius at $x_{c,i}$, $x_c$ is the radius of the mass shell 
with $M_{c,i}$ at each time, and
$M_b$ is the baryon mass in the dark matter shell at the radius $x_c$.
As a shell collapses, the baryon mass $M_b$ inside the shell increases
with the growth of $\delta_b$, namely the baryon collapsing.
Although the density fluctuation of baryons without the relative motion catches up soon with that of dark matter,
the relative motion prevents this process.
Therefore, the effect of the relative motion is included through the
evolution of $M_b$.
However it is difficult to evaluate analytically $M_b$ with the relative motion. 
Thus in order to compute Eq.~(\ref{rveq}),
we adopt $M_b$ obtained from the N-body simulation
in the following section.
We also compare the result based on the spherical collapse model with that
from the full N-body simulations in the later section.

\section{N-BODY SIMULATION}\label{C}
Besides the analytical way mentioned in the previous section,
we evaluate the effect of the supersonic motion between dark matter and
baryons on the structure formation at high redshifts by using N-body simulations.
In this section, we describe the setup of our N-body simulations.
We perform N-body simulations with the public code Gadget-2~\cite{2005MNRAS.364.1105S}.
In all N-body simulations,
the cosmological parameters are set to
$(\Omega_m,\ \Omega_\Lambda\ h)=(0.31,\ 0.69,\ 0.68)$ with
$\Omega_b/\Omega_m\sim1/6$. The effect of the supersonic relative
motion on the structure formation works after the decoupling between
photons and baryons. Therefore, the initial redshift for the simulations
is $z_i=1000$.
Note that our results almost do not depend on the cosmological
parameters because we are interested in the structure formation in the matter dominated era.

For the initial distribution of the particles,
we consider the spherical top-hat overdensity region in the isolated system.
The simulation box has the uniform distribution of the particles with a uniform overdensity sphere.
We set the box size to $L_{\rm Box}=200~{\rm kpc}/h$ and the radius of the overdensity sphere to $r_i=50~{\rm kpc}/h$.
Note that we denote hereafter $x$ as the proper distance and $r$ as the
comoving distance.
In the box, the number of the uniform particles is
$3\times 10^6$ and the initial density contrast of dark matter in
the overdense sphere is $\delta_{c,i}=0.033$.
Thus the mass of particles is $2\times 10^2~M_\odot/h$ and the mass of dark matter within the initial overdensity sphere is
given by $M_c \sim 4\times 10^7~M_\odot/h$. 
Moreover we set the softening parameter to $\epsilon=0.1~{\rm kpc}/h$.
We confirm that changes in these parameters do not affect our result
qualitatively.

According to the cosmological perturbation theory,
the amplitude of baryon density fluctuations is 1\% of that of dark matter density fluctuations
with $k=100~{\rm Mpc^{-1}}$ at $z\sim 1000$.
The initial fluctuations of baryons are negligible compared with those of
dark matter at $z_i=1000$.
Therefore, we assume that $\Omega_b/\Omega_m~(\sim1/6)$ of the uniform distributed particles
is composed of baryons and there is no baryon
fluctuation in the overdensity sphere.

\begin{figure}[tbp]
\begin{center}
	\includegraphics[width=.36\textwidth]{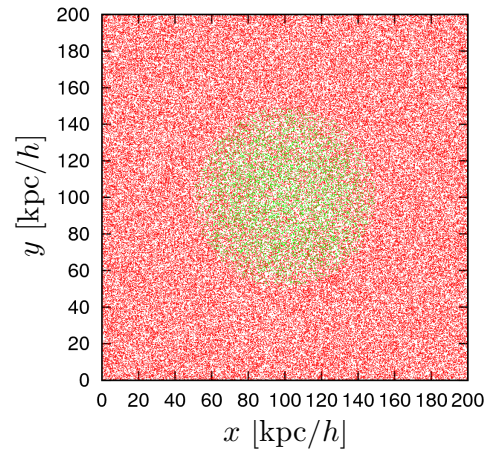}
\end{center}
  \caption{The initial configuration of the particles. 
  The green particles are contained within the top-hat sphere and the red points are otherwise.}
  \label{inipos}
\end{figure}

\begin{figure}[tbp]
\begin{center}
	\includegraphics[width=.45\textwidth]{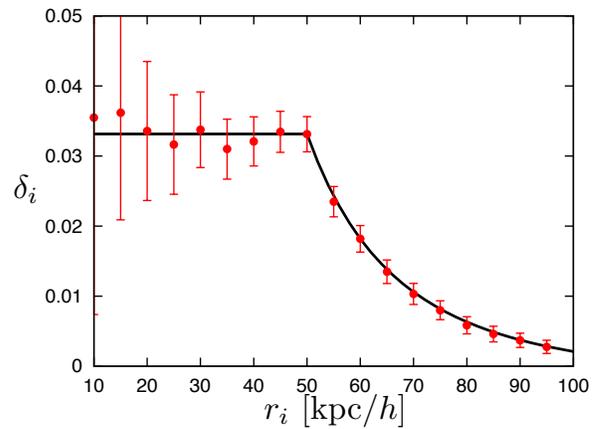}
\end{center}
  \caption{The initial density contrast distribution of dark matter. 
  The red points are the values averaged five realizations and the error bars show the shot noise caused by the number of particles contained within mass shells. 
  The black line is the analytical prediction.}
  \label{inidel}
\end{figure}

Figure~\ref{inipos} shows the initial configuration of the particles.
The red dots represent the particle uniformly distributed in the box
and the green dots are for the particles included in
the overdense sphere.
Figure~\ref{inidel} shows the initial density contrast of
dark matter particles
as a function of the radius from the center.
In this figure,
the red points indicate the values averaged over five realizations of our
simulations and the error bars represent the shot noise caused by the finite particle number.
The black line is the analytical prediction from our
initial condition.
This figure tells us that the density is constant in the top-hat sphere.
Therefore we can convert the initial position of mass shell to the mass contained within each shell by using the relation $M_{c,i}=4\pi\bar{\rho}(1+\delta_i)x_i^3/3$.
We set the initial velocity of dark matter given by 
only the second term of Eq.~(\ref{eq4}), because N-body simulations
are performed in the comoving coordinate.

In order to take into account the supersonic relative motion, we give
the additional velocity to all baryons.
The correlation of the supersonic relative velocity has the significant
value on larger scale than scales of our interest  that are smaller than Mpc. 
Therefore, we assume that all baryons in the simulations have the constant supersonic
relative velocity $v_{bc}$ in one direction.
In other words, in the simulation, the additional initial velocity for
baryons is represented as $\vec{v}_{b,i}=(v_{bc},0,0)$.

All terms related to the relative velocity in Eq.~(\ref{eqdl}) are
proportional to $k v_{bc}$. This fact suggests that the effect of
relative motion on the spherical collapse of dark matter halos also depends
on a factor $v_{bc}k\propto v_{bc}/M_c^{1/3}$.
Thus, instead of changing both the dark matter halo mass $M_c$ and the relative velocity
$v_{\rm bc}$, we perform numerical simulations for different relative velocities~($5~{\rm km}$, $15~{\rm km}$, $30~{\rm km}$,
$50~{\rm km}$, $100~{\rm km}$, $150~{\rm km}$, $200~{\rm km}$, $300~{\rm
km}$, $500~{\rm km}$) with fixing the dark matter halo mass
$M_c\sim 4\times 10^7~M_\odot/h$, in order to evaluate the dependence of
the effect of the relative velocity on $M_c$ and $v_{bc}$.  

In the simulations,
we use the periodic boundary condition.
Therefore, when the relative velocity is higher than $200~{\rm km/s}$,
baryon particles leaving the simulation box along the direction of the
relative velocity reenter
the box from the opposite direction
due to the boundary condition.
In this case, the distribution of reentering baryons is no longer
homogeneous in the perpendicular direction to the relative velocity and,
resultantly, the dependence on the boundary condition arises in the results.
In order to remove this dependence, we make the perpendicular positions
of baryons random when the baryon particles reenter the simulation box.


\section{RESULT}\label{Ca}
In this section, we present the results of our N-body simulations.
First we show the result of the reference model that is the case without the
relative velocity.
\begin{figure}[tbp]
\begin{center}
	\includegraphics[width=.45\textwidth]{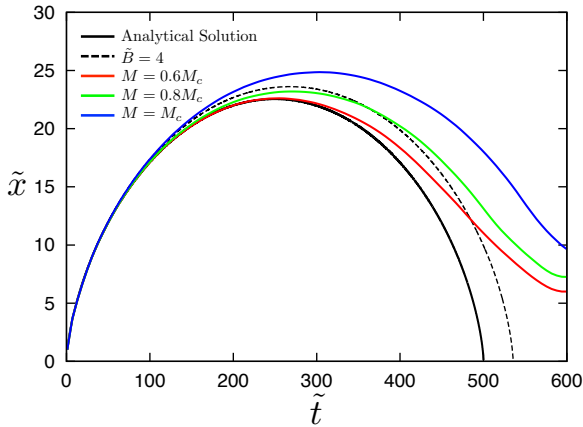}
\end{center}	
  \caption{The evolutions of radii of mass shells containing $0.6M_c$ (red), $0.8M_c$ (green), and $M_c$ (blue). 
  The solid black line is the analytical solution of spherical collapse model with $\delta_{m,i}=0.028$. 
  The dashed black line is the analytical solution corrected in consideration of the periodic boundary condition with $\tilde{B}=4$.}
  \label{sc}
\end{figure}
Figure~\ref{sc} shows the time evolutions of the radii of mass shells
from the center.
In our simulation, we determined the center of the collapsing shells by using
the mean position of the particles contained initially within the
top-hat overdensity sphere at each time step.
In this figure, the red line represents the mass shell containing $0.6
M_c\ (r_i=42~{\rm kpc/h})$, and
the green and blue lines are for that containing $0.8 M_c\ (r_i=46~{\rm
kpc/h})$ and $M_c\ (r_i=50~{\rm kpc/h})$, respectively.
Additionally the black line corresponds to the analytical solution of the spherical
collapse model, Eq.~(\ref{eq3}), with
$\delta_{m,i}=(\bar{\rho}_{c,i}\delta_{c,i}+\bar{\rho}_{b,i}\delta_{b,i})/(\bar{\rho}_{c,i}+\bar{\rho}_{b,i})=0.028$
and the velocity $v_i$ given by Eq.~(\ref{eq4}).
Note that the shell evolution of the spherical collapse model depends on
$\delta_{m,i}$ only. Therefore, in our initial condition where the overdensity
sphere has the homogeneous density profile, 
the spherical collapse model predicts that 
all mass shells inside the overdensity sphere trace the black dashed
line and collapse at the same time,
independently on the mass contained by the mass shells.

The evolutions of radii of the mass shells from N-body simulations agree with 
the analytic solution of the spherical collapse model
before the turnaround time when the radius reaches the maximum.
However, after the turnaround time, the results from N-body simulations deviate from the analytic
solution. One of the reasons for this deviation is that
the particles cannot be concentrated on the infinitesimal point in
N-body simulations.
Therefore, the particles in the simulations begin to be relaxed with each other after the turnaround
time and the collapse is prevented.
Furthermore the shot noise induces the substructures inside the
collapsing sphere and the ejection of
the particles from the mass shells, 
and resultantly causes the dispersion
of the collapse time as discussed in Ref.~\cite{2015MNRAS.446.1335W}.
These effects of relaxation and shot noise lead the delay of the collapse
and cause the deviation from the analytical solution after the turnaround time.

Figure~\ref{sc} also shows that the outer mass shells collapse later than
the inner ones, although the theoretical spherical collapse model claims that all
mass shells collapse at the same time.
The reason is the effect of the periodic boundary condition.
In order to evaluate this effect simply, we consider
the motion of a particle along the $x$-axis direction with the periodic boundary condition.
Since we should take into account the gravitational force from the overdensity region
in the other boxes due to the boundary condition,
the EoM, Eq.~(\ref{eq1}), along the $x$-axis is corrected to
\begin{align}
	\frac{d^2 \tilde{x}}{d\tilde{t}^2}=-\frac{2(1+\delta_i)}{9\tilde{x}^2}+\sum_{n=1}^\infty\left[\frac{2\delta_i}{9(na\tilde{B}-\tilde{x})^2}-\frac{2\delta_i}{9(na\tilde{B}+\tilde{x})^2}\right] \label{bcsc},
\end{align}
where $\tilde{B}=a_iL_{\rm Box}/x_i$.
When the position of the shell is close to the center of the simulation box at the
initial time, $\tilde{B}$ becomes small and vice versa. 
Namely, the smaller $\tilde{B}$ is, the more efficient the boundary effect is.
In our calculation, the most outer shell that contains the mass $M_c$
inside has $\tilde{B}=4$. 
The evolution for $\tilde{B}=4$ is plotted in the black dashed line in Fig.~\ref{sc}.

\begin{figure}[tbp]
\begin{center}
	\includegraphics[width=.45\textwidth]{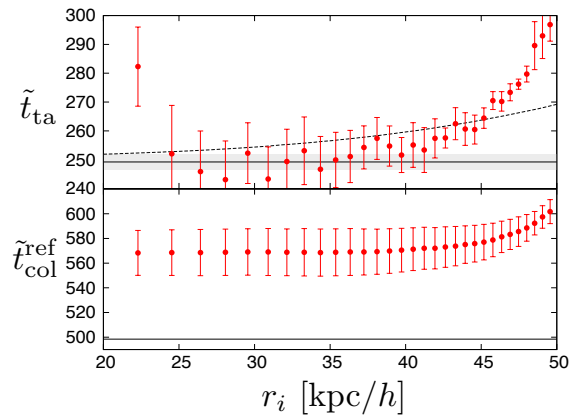}
\end{center}
  \caption{The turnaround time (upper panel) and the reference collapse time (lower panel). 
  The red points are the results of N-body simulations with the standard error measured five realizations, and the black lines are the analytical predictions. 
  In the upper panel the shaded region shows the error caused from shot noise shown in Fig.~\ref{inidel}. 
  The dashed line shows the prediction from the solutions of Eq.~(\ref{bcsc}) with $\tilde{B}$ converted from $r_i$.}
  \label{tac}
\end{figure}

Figure~\ref{tac} shows the turnaround time~(upper panel) and the reference
collapse time~(lower panel)
in the reference model as functions of the initial radius of the mass shell.
Here, the turnaround time and the reference collapse time are defined as the times
when the radius of the mass shell becomes maximum and minimum, respectively.
In this figure, the red points with the error bars represent the
averages and the standard errors obtained from the five realizations of N-body simulations, and
the black line corresponds to the theoretical predictions in the spherical collapse model.
Moreover the dashed line shows the turnaround time obtained from
Eq.~(\ref{bcsc}) which includes the effect of the boundary condition.
One can find that the turnaround time is consistent with 
the theoretical prediction within $r_i=40~{\rm kpc/}h$.
The one of the reasons why the outer mass shells turn around later is the effect of the
periodic boundary condition discussed above.
In the case where we take the boundary condition into account, the turnaround time matches the theoretical prediction within $r_i=43~{\rm kpc}/h$ that corresponds with $M=0.65M_c$. The difference between the turnaround time estimated from the outer shells than $r_i=43~{\rm kpc}/h$ and theoretical one is due to the ejections of particles.
As we have mentioned,
the reference collapse time in N-body simulations delays for all $r_i$,
compared with the theoretical prediction.
Additionally, similarly to the turnaround time, the deviation becomes
large as $r_i$ increases.

\begin{figure}[tbp]
\begin{center}
	\includegraphics[width=.45\textwidth]{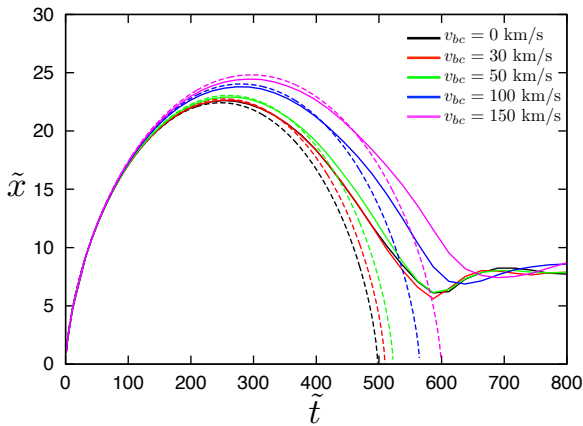}
\end{center}
  \caption{The evolutions of mass shell containing $0.6 M_c$ with supersonic relative velocities $v_{bc}=30~{\rm km}$ (red), $v_{bc}=50~{\rm km}$ (green),  $v_{bc}=100~{\rm km}$ (blue), $v_{bc}=150~{\rm km/s}$ (magenta) and reference model (black). 
  Each dashed line is the solution of Eq.~(\ref{rveq}) with the baryon density fluctuation derived from N-body simulations.}
  \label{rvsc}
\end{figure}

Next we show how the supersonic relative motion affects the collapse in N-body simulations.
Performing five realizations for different $v_{bc}$, we
obtain the typical evolution of mass shells by averaging each realization.
Figure~\ref{rvsc} represents the evolutions of the mass shells containing
$0.6M_c$ with four different supersonic relative velocities,
$v_{bc}=30~{\rm km/s}$~(in red), $v_{bc}=50~{\rm km/s}$~(in green),
$v_{bc}=100~{\rm km/s}$~(in blue) and $v_{bc}=150~{\rm km/s}$~(in
magenta).
We can convert the results for different
velocities with a fixed mass into those for different masses with a
fixed velocity
through the dependence of the relative velocity effect on
$v_{bc}/M_c^{1/3}$ as mentioned above.

For comparison, the corresponding evolution of the mass shell with $0.6M_c$
in the reference model~($v_{bc}=0~{\rm km/s}$) is plotted as the black solid line.
As the relative velocity becomes large, the start of the collapse delays
and the maximum radius increases.
Therefore, we can conclude that the supersonic relative motion prevents the collapse.
\begin{figure}[tbp]
\begin{center}
	\includegraphics[width=.45\textwidth]{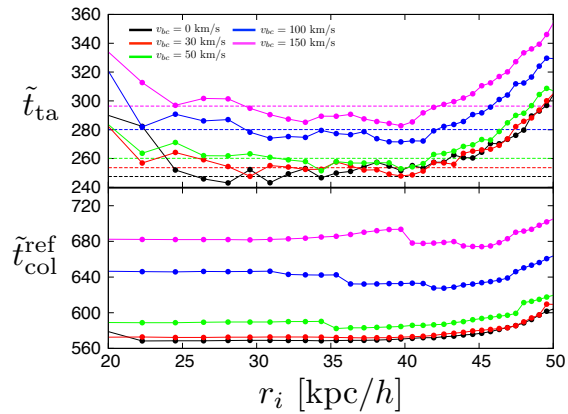}
\end{center}
  \caption{The turnaround time (upper panel) and the reference collapse time (lower panel) as the function of the initial radius of the mass shell with the supersonic relative velocity $v_{bc}=30~{\rm km}$ (red), $v_{bc}=50~{\rm km}$ (green),  $v_{bc}=100~{\rm km}$ (blue), $v_{bc}=150~{\rm km/s}$ (magenta) and the reference model (black). 
  In the upper panel the dashed lines are the turnaround time estimated from the solution of the semianalytical model.}
  \label{da}
\end{figure}
Additionally we show the solutions of Eq.~(\ref{rveq}) as the dashed lines in Fig.~\ref{rvsc}.
Solving Eq.~(\ref{rveq}) numerically, we use the baryon fluctuation
$\delta_b$ obtained from the particle data of the
N-body simulations within $r_i\leq 42~{\rm kpc/h}$ corresponding to $0.6M_c$.
We find that the semianalytical model agrees with the results of N-body simulations before the turnaround time.

To illustrate
the delay of the collapse due to the supersonic relative motion 
we plot the turnaround time and the reference collapse time in Fig.~\ref{da}
for the different supersonic relative velocities.	
In this figure, both the turnaround and the reference collapse time are represented 
as the functions of the initial radius of the mass shell.
We show additionally the turnaround times obtained from the solutions of Eq.~(\ref{rveq}) as the dashed lines in the top panel of Fig.~\ref{da}.
The evaluations of the turnaround time from the semianalytical solutions
are consistent with the turnaround times from N-body simulations within
$r_i\sim40~{\rm kpc}/h$ that is same as the reference case.
In the following discussions,
we use twice the turnaround time as the collapse time.
In this case, the scale factor at the collapse time is given by
$a_{\rm col}=\left(2\tilde{t}_{\rm ta}\right)^{2/3}a_i$.

\section{DISCUSSION}\label{D}
In this section, we show the baryon fraction within the dark matter
overdensity sphere. We also discuss the delay of the collapse time
and provide the fitting formula of the critical density contrast with the relative motion between dark matter and baryons.
Furthermore we present the modification of the halo mass function by taking account of the relative motion.

\subsection{Baryon fraction}\label{Db}
First we consider the baryon fraction which represents the mass ratio
between baryons and total matter~($f_b=M_b/M_m$) 
within the dark matter over-density sphere.
The baryon fraction is important not only to estimate the effect of
the supersonic relative motion on the collapse of dark matter spheres,
but also to discuss the first star formation or observables related
with baryons.
We calculate the baryon fraction by counting baryon particles within the dark
matter collapsing over-density sphere.

\begin{figure}[tbp]
\begin{center}
	\includegraphics[width=.45\textwidth]{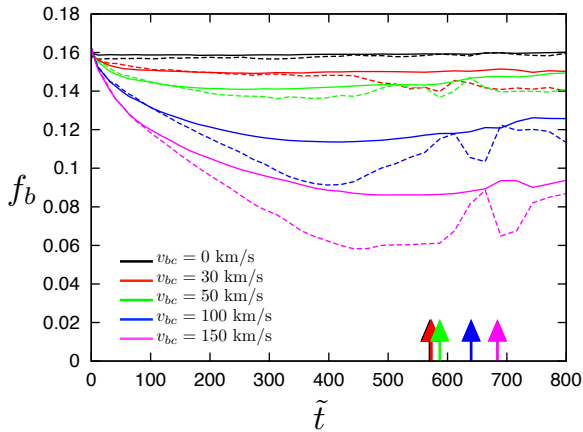}
\end{center}
  \caption{The baryon fractions within the dark matter over-density sphere whose initial radius are $50~{\rm kpc}/h$ (solid lines) and $42~{\rm kpc}/h$ (dashed lines). 
  The arrows show the reference collapse times.}
  \label{bf}
\end{figure}

Figure~\ref{bf} shows the time evolutions of the baryon fraction. 
As the relative velocity increases,
the baryon fraction becomes smaller.
In the case with the nonzero relative velocity, the baryon overdensity
region is no longer spherically symmetric and the peak position of baryon density is different
from that of dark matter, depending on the amplitude of the relative velocity.
However, such asymmetry of the baryon distribution does not affect the dark matter collapse well.
As shown in Fig.~\ref{rvsc}, the dark matter collapse in the
N-body simulations is consistent with the spherical collapse model until the turnaround time.
Therefore, we can infer that, in spite of the asymmetric distribution
for baryons,
the dark matter collapse remains spherical.  
Around the reference collapse time of the dark matter shells, the baryon fraction estimated within $r_i=42~{\rm kpc}/h$ starts to oscillate. 
This is mainly due to the difference of the density peak positions
between baryons and dark matter. The baryon overdensity region is attracted
by that of dark matter gravitationally and oscillates around.
As time goes, the difference of the peak positions
will be relaxed and the peak position of baryons is expected to overlap
that of dark matter. 
Note that this result is based on the spherical collapse model which
is an ideal isolated
system. However, the actual collapse happens with many surrounding
effects as shown in cosmological simulations. Therefore, to evaluate the baryon
fraction properly, these effects could be not negligible.

\subsection{Delay of the halo formation}\label{Dc}
The supersonic relative motion delays the collapse time
of dark matter halos as shown in Fig.~\ref{da}.
In the spherical collapse model without the relative velocity,
the collapse time is dependent on only the initial density fluctuation.
However, in the case of the nonzero relative velocity,
the collapse time depends on the halo mass $M_c$ and the amplitude of the relative velocity. 
Additionally, the effect of the relative motion cumulatively becomes large for
the small initial density contrast, because the small initial density contrast takes longer time to the collapse.
Therefore, the change of the collapse time with the relative velocity
is represented as a function of $M_c$, $v_{bc}$ and $\delta_{c,i}$
We define the correction of the scale factor at the
collapse time related with the modification of the critical density contrast, as
\begin{align}
	\mathcal{A}(M_c,v_{bc},\delta_{c,i})&\equiv\frac{a_{\rm
 col}(M_c,v_{bc},\delta_{c,i})-a_0(\delta_{c,i})}{a_0(\delta_{c,i})},\label{eq:col}
\end{align}
where $a_0(\delta_{c,i})$ is the scale factor at the collapse time without the relative velocity between baryons and dark matter.

\begin{figure}[tbp]
\begin{center}
	\includegraphics[width=.45\textwidth]{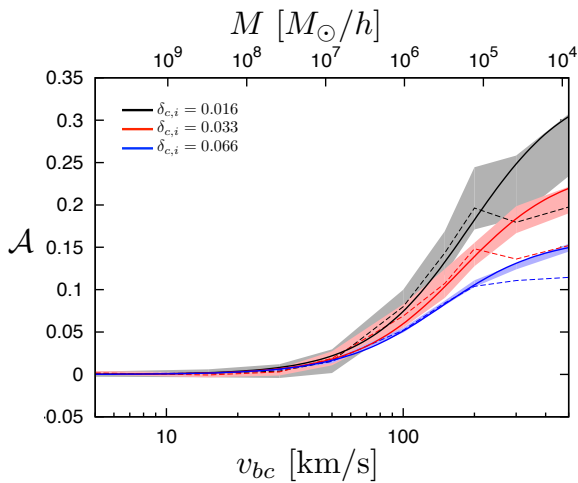}
\end{center}
  \caption{The relative time difference for the collapse as a function of relative velocity $v_{bc}$ with the halo mass fixed to $M_c\sim 4\times
10^7~M_\odot/h$ and three initial density fluctuations $\delta_{c,i}=0.016$ (black), $\delta_{c,i}=0.033$ (rad) and $\delta_{c,i}=0.066$ (blue) evaluated from turnaround time. 
The shaded regions show the standard error region from the N-body simulation. 
The solid lines are the fitting formula Eq.~(\ref{ff}). 
The dashed line are results from the N-body simulations with the usual periodic boundary condition. 
The upper horizontal axis shows the mass converted with $v_{bc}=30~{\rm km/s}$.}
  \label{rda}
\end{figure}

Figure~\ref{rda} shows the relative difference~$\mathcal {A}$
as a function of the relative velocity with a fixed mass~$M_c\sim 4\times 10^7~M_\odot/h$.
We plot $\cal A$ for different three initial density contrasts estimated from the N-body 
simulations with our boundary condition (shaded regions) or the usual periodic boundary condition 
(dashed lines).
We find that the effect of the supersonic relative velocity is negligible for velocities smaller than 50~km/s.
The relative velocity is small so that baryons are captured
gravitationally by the dark matter halo and accrete to the
halo. Therefore, we can roughly estimate the threshold velocity as the
circular velocity of the dark matter halo at the initial redshift,
\begin{align}
	v_{\rm cir}=\sqrt{\frac{GM_c}{x_i}}\simeq 57~\left(\frac{M_c}{3.85\times 10^7~M_\odot}\right)^{1/3}~{\rm km/s}.
\end{align}
This criterion can be also obtained from the condition that the third term related to the relative velocity, 
$\mu^2 k^2$, dominates in the right-hand side in the second equation of Eq.~(\ref{eqdl}). 
Therefore, when the relative velocity is larger than the criterion velocity, the relative motion prevents the collapse by the third term in Eq.~(\ref{eqdl}).
Similarly, when the relative velocity is larger than $v_{\rm cir}$, the
effect of the relative motion on the structure formation arises.
The upper horizontal axis represents the corresponding mass in the
case of a fixed velocity~$v_{bc} = 30~{\rm km/s}$, which is converted
through the $v_{bc}/M_c^{1/3}$-dependence of the relative motion effect.
Thus one can find that the supersonic relative motion does not affect
the formation of dark matter halos with mass larger than $M_c\simgt 10^7~M_\odot/h$ for $v_{bc}=30{\rm km/s}$
which corresponds to the effective Jeans scale discussed in Sec.~\ref{Ba}.

In the N-body simulation results with the usual periodic boundary condition,
the modification becomes independent of the amplitude of the relative
velocity in the case with $v_{bc}\simgt 200~{\rm km/s}$ shown in Fig.~\ref{rda}.
However this independence is due to the artificial condition of the simulations.
When $v_{bc}\simgt 200~{\rm km/s}$, all baryons can travel a
distance larger than the simulation box size $L_{\rm Box}$ until $\tilde{t}\sim100$
and are attracted gravitationally twice by the collapsing dark matter sphere.
Therefore, the effect of the relative velocity seems to be
saturated in this relative velocity region. 
On the other hand, the results of the N-body simulations with our boundary condition 
shown as the shaded region in Fig.~\ref{rda} are not saturated.
In order to verify the validity of the result of our N-body simulations,
we show the collapse time and 
the baryon fraction in the limit of the large homogeneous relative velocity.
One can easily imagine that, 
when the relative velocity is enough high~($v_{bc}=500~{\rm km/s}$),
which corresponds to very small dark matter halos ($M_c\sim10^4
~M_\odot/h$) in the case of $v_{bc}=30~{\rm km/s}$,
baryons do not collapse along the relative velocity direction.
In this limit, we can ignore
the gravitational force from the dark matter halo on the baryon motion along
the relative velocity direction.
In other words, the gravitational collapsing of baryons occurs
perpendicular to the relative velocity direction and does not along the
parallel direction.
Therefore, to evaluate the evolution of the density fluctuations,
it is useful to consider the motion of baryons with the relative
velocity in the comoving cylindrical coordinate system whose axis is parallel to
the relative velocity motion.
In this case, 
the EoM of baryons particles with the initial
velocity~$(v_{b\parallel},v_{\perp})=(v_{bc},0)$ is given as
\begin{align}
	\frac{d^2\vec{r}_b}{dt^2}=&-2H\vec{v}_b-\frac{G\delta M_c}{a^3r_b^3}\vec{r}_b,\notag\\
	\delta M_c=&\frac{4}{3}\pi \bar{\rho}_c[r_{c,i}^3(1+\delta_{c,i})-r_c^3]\notag\\
	&\times \max[1,(r_b/r_c)^3],\label{lim}
\end{align}
where $r_b$ is the radial component of $\vec{r}_b$ from the center of the dark matter sphere and $r_c$ is the radius of the overdensity sphere of dark matter.
Note that we ignore the gravitational force of the baryon fluctuation in Eq.~(\ref{lim}), because $\delta M_b\ll \delta M_c$.
We solve Eqs.~(\ref{rveq}) and (\ref{lim}) numerically with different
initial positions $\vec{r}_{b,i}$ chosen randomly.
We calculate the resultant baryon density contrast in a collapsing
spherical shell of dark matter by taking the average of 
$\delta_b=(r_{bi\perp}/r_{b\perp})^{2}-1$ for baryons inside the shell of dark matter,
 because the collapse of baryons is cylindrical.

\begin{figure}[tbp]
\begin{center}
		\includegraphics[width=.45\textwidth]{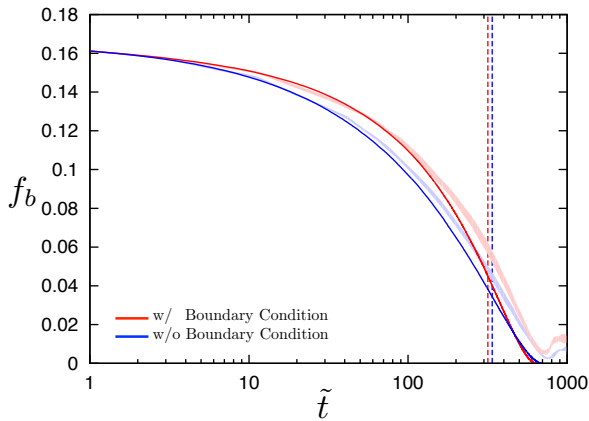}
\end{center}
  \caption{The baryon fractions within the dark matter halo ($r_i=42~{\rm kpc}/h$) with $M_c\sim 4\times 10^7 M_\odot/h$ and $v_{bc}=500~{\rm km/s}$ and solution of Eqs.~(\ref{rveq}) and (\ref{lim}) with or without the periodic boundary condition (red or blue line). 
  The each color shaded region show the 1$\sigma$ dispersion of the baryon fractions estimated from five realizations of N-body simulations. 
  The vertical dashed lines show the turnaround times from N-body simulations.}
  \label{del}
\end{figure}

Figure~\ref{del} shows the evolution of the baryon fractions with $M_c\sim 4\times 10^7 M_\odot/h$ and $v_{bc}=500~{\rm km/s}$.
The red solid line represents the solution obtained with the
periodic boundary condition.
In order to take into account the periodic boundary condition, we
solve Eqs.~(\ref{rveq}) and (\ref{lim}) with the assumption
that the position of baryons, $\vec{r}_b$, 
is limited within the box size and baryons return into the box from the
opposite side when they exit from one side of the box.
For comparison, 
we also solve the equations without the boundary condition and plot the
solution in the blue solid line.
Baryons with the periodic boundary condition feel gravitational force stronger than without the
boundary condition.
Therefore, the collapse is faster with the boundary condition than
without the boundary condition.

Moreover, in Fig.~\ref{del},
we show the results from N-body simulations.
The red shaded region represents the standard error region from the
N-body simulations with the periodic boundary condition, while the blue
shaded region gives the standard error region for the N-body simulations with our
boundary condition which is the periodic boundary condition with the position
shuffling of the baryon particles reentering into the box.
In Fig.~\ref{del}, N-body simulations with our boundary condition is
consistent with the numerically solution without the periodic boundary
condition, while N-body simulations with the periodic boundary
condition agrees with the solution with the periodic boundary condition.
We find that, in the both boundary condition cases, the differences
between the numerical solutions and N-body simulations arise around  
$\tilde{t}\simeq 100$. 
This is because, as the collapse proceeds, the baryon density in the
N-body simulations grows as the spherical collapse 
rather than the cylindrical one. Therefore, the baryon fraction is
larger in N-body simulations than in the numerical calculations.

\begin{figure}[tbp]
\begin{center}
	\includegraphics[width=.45\textwidth]{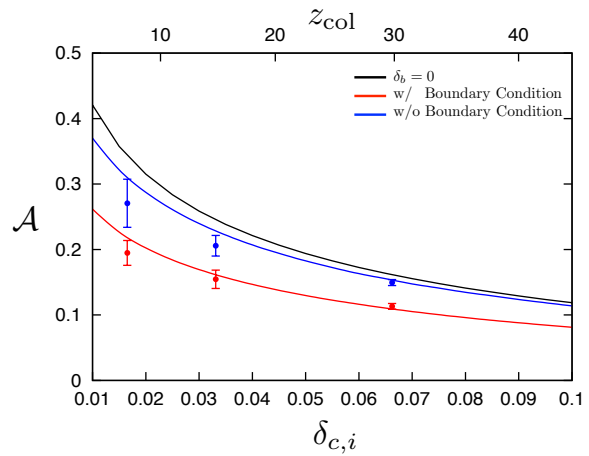}
\end{center}
  \caption{The collapse time with $M_c\sim 4\times 10^7 M_\odot/h$ and $v_{bc}=500~{\rm km/s}$ estimated by solving Eqs.~(\ref{rveq}) and (\ref{lim}) with boundary condition (red) and without boundary condition (blue). 
  The black solid line is corresponded the solution of Eqs.~(\ref{rveq}) with $\delta_b=0$ at any time. 
  Points represent estimations from N-body simulations with $v_{bc}=500~{\rm km/s}$. 
  The upper axis presents the collapse time from Eq.~(\ref{eq3}) with the total matter density fluctuation $\delta_{m,i}=\bar{\rho}_c\delta_{c,i}/(\bar{\rho}_c+\bar{\rho}_b)$.}
  \label{limda}
\end{figure}

In addition, we plot $\mathcal{A}$ as functions of the initial density fluctuations
$\delta_{c,i}$ with $M_c\sim 4\times 10^7 M_\odot/h$ and $v_{bc} =500~{\rm km/s}$
in Fig.~\ref{limda}. 
The red and blue solid lines are obtained from the numerical calculations
with and without the
periodic boundary condition, respectively.
In the case with the periodic boundary condition, we overestimate the
gravitational force to collapse as mentioned above and, therefore, the
delay of the collapse is not larger than in the case without the boundary condition.
For comparison, we plot the black solid line which represents the results with
the assumption that baryons cannot collapse.
The difference from the black solid line represents the contribution due the collapse of the baryon component. 
When the relative velocity is large enough, baryons cannot collapse to
the dark matter mass shell. Accordingly, as the relative velocity
becomes large, the blue solid line
shifts to the black line. 
We also plot the results of N-body simulations with the periodic boundary condition
and our boundary condition as red and blue points with the standard
error bars in Fig.~\ref{limda}, respectively.
As shown in Fig.~\ref{del}, the numerical calculation with the periodic
boundary condition agrees with N-body simulations with the periodic
boundary condition, while the numerical calculation without the boundary
condition is consistent with N-body simulations with our boundary
condition.
We remind you that our boundary condition is introduced to remove 
the artificial distribution of the reentering baryon particles
due to the periodic boundary condition in N-body simulations.
We conclude that the saturation in $\mathcal{A}$ from N-body simulations with the periodic boundary condition
is caused by this artifact.
The results from N-body simulations with our boundary condition present realistic phenomena.

Based on the results of our N-body simulations,
we find the fitting formula of $\cal A$ represented as the solid lines
in Fig.~\ref{rda}.
The fitting formula is given by
\begin{align}
	\mathcal{A}(M_c,v_{bc},\delta_{c,i})&=\mathcal{A}_{\delta_b=0}(\delta_{c,i})\frac{\mathcal{B}^\nu(M_c,v_{bc},\delta_{c,i})}{\mathcal{B}^\nu(M_c,v_{bc},\delta_{c,i})+1},\notag\\
	\mathcal{B}(M_c,v_{bc},\delta_{c,i})&=\frac{v_{bc}}{v_{\rm norm}(\delta_{c,i})}\left(\frac{M_c}{3.85\times 10^7~M_\odot/h}\right)^{-1/3},\notag\\
	v_{\rm norm}(\delta_{c,i})&=a_v - b_v\delta_{c,i},\label{ff}
\end{align}
where 
$\mathcal{A}_{\delta_b=0}$ is the solution of Eq.~(\ref{rveq}) with $\delta_b=0$
shown in Fig.~\ref{limda},
and
$\nu$, $a_v$ and $b_v>0$ are the fitting parameters.
The velocity $v_{\rm norm}$ is the critical velocity for the
collapse of baryons along the direction of the relative velocity.
When $v_{\rm bc} \gg v_{\rm norm}$,
the relative velocity is much larger than $v_{\rm circ}$ even at the
collapse time and baryons does not collapse along the direction of the relative velocity
as mentioned above. Since the collapse time becomes long with
decreasing $\delta_i$ and $v_{\rm bc}$ is inversely proportional to the
scale factor, $v_{\rm norm}$ increases as $\delta_i$ decreases.
Thus we conclude that the delay is controlled by two critical velocity $v_{\rm circ}$ and $v_{\rm norm}$.
Nevertheless it can be calculated numerically, we use the approximated
function of $\mathcal{A}_{\delta_b=0}$, 
\begin{align}
	\mathcal{A}_{\delta_b=0}(\delta_{c,i})=\frac{\delta_{c,i}^{-0.146}-1.06}{4.12\times \delta_{c,i}^{0.648}+1.93}.
\end{align}
We estimate the parameters by fitting simultaneously Eq.~(\ref{ff}) to
$\mathcal{A}$ with using three different initial density fluctuations.
Furthermore we perform a Fisher analysis and obtain the parameters as
\begin{align}
	\nu=2.02\pm0.07,\ a_v=205\pm16,\ b_v=877\pm253,
\end{align}
where the standard errors are estimated after marginalizing over other
parameters. Since we sample the data for three different initial conditions,
the parameter $b_v$ has a large error. However, we find that the form in Eq.~(\ref{ff}) 
fits well with the N-body simulation results. 

\subsection{Mass function}
The delay of the halo formation due to the relative velocity modifies
the abundance of dark matter halos. In this section, we evaluate the
modification based on the Press-Schechter formalism.
In the Press-Schechter formalism, the delay of the collapse is
represented as the increase of the critical density contrast.
Using the relative time difference $\cal A$, we can write the modified
critical density contrast during the matter dominated era as
\begin{align}
	\tilde{\delta}_{\rm crit}(M_c,v_{bc},\delta_{c,i})&=\delta_{\rm crit}\left[1+\mathcal {A}(M_c,v_{bc},\delta_{c,i})\right].\label{eq:delta}
\end{align}
The critical density contrast depends on the relative velocity in
the region where the collapses happens.
Therefore, the modified halo mass distribution can be written with the probability distribution function of the amplitude of the
relative velocity at the initial time $f(v_{bc})$ as
\begin{align}
	\tilde{n}(M_c,z)=&\int dv_{bc}~f(v_{bc})\sqrt{\frac{2}{\pi}}\frac{\bar{\rho}(z)}{M_c}\frac{\tilde{\delta}_{\rm crit}(M_c,v_{bc},\delta_{c,i})}{\sigma(M_c,z)}\notag\\
	&\times\left[\frac{d\ln\tilde{\delta}_{\rm crit}(M_c,v_{bc},\delta_{c,i})}{dM_c}-\frac{d\ln\sigma(M_c,z)}{dM_c}\right]\notag\\
	&\times\exp\left[-\frac{\tilde{\delta}_{\rm
 crit}^2(M_c,v_{bc},\delta_{c,i})}{2\sigma^2(M_c,z)}\right]. \label{ps}
\end{align}
Note that, because of the existence of the relative motion, the redshift
of the collapse time depends not only on the initial density fluctuation
of cold dark matter, but also on the halo mass $M_c$ and the amplitude of
relative velocity $v_{bc}$. Therefore, in Eq.~(\ref{ps}), 
the mass derivative term of the critical density additionally arises
because the mass dependence of the critical density contrast
affects the hierarchical structure formation. 

We assume that the probability distribution $f(v_{bc})$ 
follows the Maxwell-Boltzmann distribution because the each component
of relative velocity is independent and obey
the same Gaussian distribution whose mean value is zero 
and the dispersion is $\sigma_v$;
\begin{align}
	f(v_{bc}) dv_{bc} =4\pi v_{bc}^2
 \left(\frac{3}{2\pi\sigma_v^2}\right)^{3/2}\exp\left(-\frac{3v_{bc}^2}{2\sigma_v^2}\right)
 dv_{bc},
\end{align}
where we use $\sigma_v=28.8~{\rm km/s}$ according to the latest cosmological
parameters from PLANCK paper \cite{2014A&A...571A..16P} and we use CAMB\cite{2000ApJ...538..473L} to calculate $\sigma(M_c,z)$.
\begin{figure}[tbp]
\begin{center}
	\includegraphics[width=.45\textwidth]{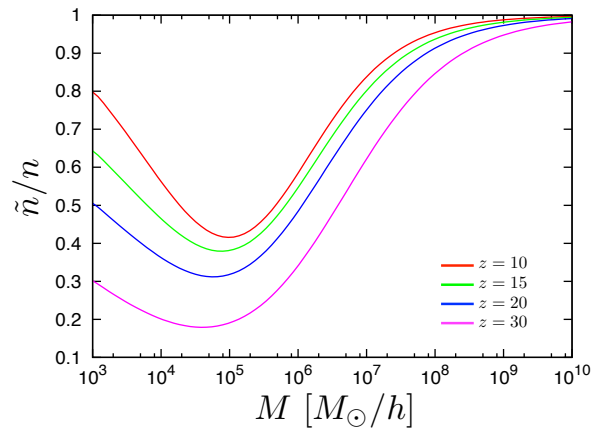}
\end{center}
  \caption{The ratio of mass function of the dark matter halo between with and without relative motion at four redshifts $z=10$ (red), $z=15$ (green) $z=20$ (blue) and $z=30$ (magenta).}
  \label{mf}
\end{figure}

Figure~\ref{mf} shows the ratio between the mass function with and
without the relative velocity.
Here we plot the ratio at four different redshifts, $z=$10, 15, 20 and 30.
These lines are evaluated by using the fitting formula Eqs.~(\ref{ff})
for $\mathcal{A}$ in Eq.~(\ref{eq:delta}).
The suppression of the mass function due to the relative motion is more
significant for smaller masses region and at higher redshifts.
At $z = 30$, although the modification $\mathcal{A}$ is
very small around $10^7 M_\odot/h\lesssim M_c \lesssim 10^9~M_\odot/h$, the
suppression of the mass function is not negligible. This is because such
massive halos at high redshifts are rare objects which satisfy
$\tilde{\delta}_{\rm crit}\gg \sigma(M_c)$ in the Press-Schechter formalism. Therefore, the
effect of the modification $\mathcal{A}$ appears exponentially in the
Press-Schechter formalism, even if
$\mathcal{A}$ is small.
The ratio of the mass functions 
reaches the minimum at $M_c\sim
10^5~M_\odot/h$. 
On smaller scales than $M_c\sim
10^5~M_\odot/h$, the suppression of the mass function decreases because
the mass derivative term of the critical density in Eq.~(\ref{ps}) increases around $M_c\sim 10^5~M_\odot/h$ as shown in Fig.~\ref{rda}.

Figure~\ref{mf} also shows that the mass function is suppressed even around
the EoR~($z\sim 10$). This is because the redshift dependence of the derivative terms in
Eq.~(\ref{ps}) is weak.
Although the suppression scales are consistent, the suppression of mass
function around EoR is stronger than in other previous works. Especially 
Ref.~\cite{2012ApJ...747..128N} reported that the suppression disappears
at $z \sim 10$ in their SPH simulations.
Of course, unlike the SPH simulations,
we only simulate the gravitational force, ignoring the baryon physics.
However, it is worth mentioning reasons of the difference between our calculation of the mass
 function and previous work in term of the gravitational growth.
The first reason for this difference is that we do not include the environmental
effects of the structure formation,~e.g., an accretion of other density
peaks of baryons neighboring the dark matter halo.
These effects increase the baryon fraction within a dark matter halo and promotes the halo formation. 
The second reason is that the relative motion produces halos derived from offsetting baryon peaks \cite{2014ApJ...791L...8N}.
This effect increases simply the number of halos. 
It is difficult to include this effect in the Press-Schechter formalism.
We will be able to estimate a number of halos originated from the baryon peaks by applying the our simulations or our semianalytical model of Eq.~(\ref{rveq}).
Finally, the initial condition for matter density and velocity fields is
still debatable in the  numerical simulations with the relative velocity.
In our simulation, we use the initial condition that leads to the maximum delay of the collapse time.
Thus the halo mass function in Fig.~\ref{mf} is calculated on the basis of the optimistic case 
where the baryons can escape most efficient from the dark matter halo.

\section{SUMMARY}\label{E}
The relative motion between baryons and dark matter plays an important
role, particularly, in small-scale structure formation at high
redshifts. We have studied their effect on the dark mater halo formation.

We have evaluated the delay of the dark matter halo collapse due to the
supersonic relative motion by using the cosmological N-body simulation.
We have found that the delay of the collapse becomes large for a dark
matter halo with $M_c\sim4\times 10^7~M_\odot/h$, 
when the relative velocity is larger than $v_{\rm cir}=57~{\rm km/s}$. 
In other words, the delay of the collapse happens when the relative
velocity is larger than
the typical circular velocity of the dark
matter halo. 
Moreover, we have shown that the supersonic relative motion delays
the fall of baryons into the potential well of the dark matter halo in the
context of the spherical collapse model.
We have also evaluated the baryon fraction $M_b/M_m$ of the dark matter
halos with the supersonic relative motion by the N-body simulation. 
The baryon fraction becomes smaller as the amplitude of the
relative motion increases.
We have pointed out that, when the relative velocity is large enough to
escape from the potential of dark matter halos, baryons can collapse
only along the perpendicular direction of the relative velocity, like
the cylindrical collapse. 
Furthermore we show the delay of the collapse time for dark matter halo by the relative motion 
depends on the initial density fluctuation within dark matter spheres, which 
determines the collapse time of the dark matter halo without relative motion. 
The smaller initial density fluctuation lead the longer time during which the supersonic relative 
motions affect the halo collapse.
In consequence, the effect of the relative motion is more efficient on dark matter halos formed at later time.
 
Finally we have estimated the suppression of the abundance of dark matter
halos by supersonic relative motion in the context of the spherical collapse model.
In the Press-Schechter formalism, the delay of the collapse increases the
critical density contrast for the collapse. We have found the fitting
formula of the critical density contrast depending on the halo mass, the
initial density fluctuations and, the relative velocity. 
Using the fitting formula, we have calculated the mass function of dark
matter halos. The relative motion decreases the mass function with mass
smaller than $10^{8}~M_\odot /h$ before EoR. 
In particular, the abundance of halos with $M_c=10^5~M_\odot/h$ is suppressed by 80\% at $z=30$ and a half at $z=10$.

The delay of the dark matter halo collapse and the decrease of the baryon
fraction in dark matter halos due to the relative motion can give
the effect on first star formation and the reionization history~\cite{2011ApJ...736..147G,2011MNRAS.412L..40M,2012MNRAS.424.1335F}.
Such effect could impact the cosmological signals of the EoR
including the CMB polarization~\cite{2012PhRvD..85d3523F}, the redshifted
21~cm lines~\cite{2011arXiv1110.4659B,2012ApJ...760....3M} 
and these cross-correlation~\cite{2008MNRAS.389..469T}.
Moreover, the relative motion between dark matter and baryons influences
the large scale structure, e.g., baryon acoustic
oscillation~\cite{2011JCAP...07..018Y,2015MNRAS.448....9S,2013PhRvD..88j3520Y},
and there is the challenging work detecting this
effect by using the results of galaxy distribution from two independent
galaxy spectroscopic survey \cite{2015arXiv150603900B}.
Based on the results of this paper, we will investigate the
effect of the relative motion on the cosmological signals probed by
ongoing or planned observations.

\begin{acknowledgements}
This work was supported by JSPS KAKENHI Grant No. 26-2667 (S.A.), No. 24340048 (K.I.) and No. 15K17646 (H.T.). H.T. also acknowledges the support by MEXT's Program for Leading Graduate Schools PhD professional, ``Gateway to Success in Frontier Asia''.
\end{acknowledgements}

\bibliography{rvsc_v2}

\begin{thebibliography}{33}
\expandafter\ifx\csname natexlab\endcsname\relax\def\natexlab#1{#1}\fi
\expandafter\ifx\csname bibnamefont\endcsname\relax
  \def\bibnamefont#1{#1}\fi
\expandafter\ifx\csname bibfnamefont\endcsname\relax
  \def\bibfnamefont#1{#1}\fi
\expandafter\ifx\csname citenamefont\endcsname\relax
  \def\citenamefont#1{#1}\fi
\expandafter\ifx\csname url\endcsname\relax
  \def\url#1{\texttt{#1}}\fi
\expandafter\ifx\csname urlprefix\endcsname\relax\def\urlprefix{URL }\fi
\providecommand{\bibinfo}[2]{#2}
\providecommand{\eprint}[2][]{\url{#2}}

\bibitem[{\citenamefont{{Aubourg} et~al.}(2014)\citenamefont{{Aubourg},
  {Bailey}, {Bautista}, {Beutler}, {Bhardwaj}, {Bizyaev}, {Blanton},
  {Blomqvist}, {Bolton}, {Bovy} et~al.}}]{2014arXiv1411.1074A}
\bibinfo{author}{\bibfnamefont{{\'E}.}~\bibnamefont{{Aubourg}}},
  \bibinfo{author}{\bibfnamefont{S.}~\bibnamefont{{Bailey}}},
  \bibinfo{author}{\bibfnamefont{J.~E.} \bibnamefont{{Bautista}}},
  \bibinfo{author}{\bibfnamefont{F.}~\bibnamefont{{Beutler}}},
  \bibinfo{author}{\bibfnamefont{V.}~\bibnamefont{{Bhardwaj}}},
  \bibinfo{author}{\bibfnamefont{D.}~\bibnamefont{{Bizyaev}}},
  \bibinfo{author}{\bibfnamefont{M.}~\bibnamefont{{Blanton}}},
  \bibinfo{author}{\bibfnamefont{M.}~\bibnamefont{{Blomqvist}}},
  \bibinfo{author}{\bibfnamefont{A.~S.} \bibnamefont{{Bolton}}},
  \bibinfo{author}{\bibfnamefont{J.}~\bibnamefont{{Bovy}}},
  \bibnamefont{et~al.}, \bibinfo{journal}{ArXiv e-prints}
  (\bibinfo{year}{2014}), \eprint{1411.1074}.

\bibitem[{\citenamefont{{Planck Collaboration}
  et~al.}(2015)\citenamefont{{Planck Collaboration}, {Adam}, {Ade}, {Aghanim},
  {Akrami}, {Alves}, {Arnaud}, {Arroja}, {Aumont}, {Baccigalupi}
  et~al.}}]{2015arXiv150201582P}
\bibinfo{author}{\bibnamefont{{Planck Collaboration}}},
  \bibinfo{author}{\bibfnamefont{R.}~\bibnamefont{{Adam}}},
  \bibinfo{author}{\bibfnamefont{P.~A.~R.} \bibnamefont{{Ade}}},
  \bibinfo{author}{\bibfnamefont{N.}~\bibnamefont{{Aghanim}}},
  \bibinfo{author}{\bibfnamefont{Y.}~\bibnamefont{{Akrami}}},
  \bibinfo{author}{\bibfnamefont{M.~I.~R.} \bibnamefont{{Alves}}},
  \bibinfo{author}{\bibfnamefont{M.}~\bibnamefont{{Arnaud}}},
  \bibinfo{author}{\bibfnamefont{F.}~\bibnamefont{{Arroja}}},
  \bibinfo{author}{\bibfnamefont{J.}~\bibnamefont{{Aumont}}},
  \bibinfo{author}{\bibfnamefont{C.}~\bibnamefont{{Baccigalupi}}},
  \bibnamefont{et~al.}, \bibinfo{journal}{ArXiv e-prints}
  (\bibinfo{year}{2015}), \eprint{1502.01582}.

\bibitem[{\citenamefont{{Furlanetto} et~al.}(2006)\citenamefont{{Furlanetto},
  {Oh}, and {Briggs}}}]{2006PhR...433..181F}
\bibinfo{author}{\bibfnamefont{S.~R.} \bibnamefont{{Furlanetto}}},
  \bibinfo{author}{\bibfnamefont{S.~P.} \bibnamefont{{Oh}}}, \bibnamefont{and}
  \bibinfo{author}{\bibfnamefont{F.~H.} \bibnamefont{{Briggs}}},
  \bibinfo{journal}{\physrep} \textbf{\bibinfo{volume}{433}},
  \bibinfo{pages}{181} (\bibinfo{year}{2006}), \eprint{astro-ph/0608032}.

\bibitem[{\citenamefont{{Pritchard} and {Loeb}}(2012)}]{2012RPPh...75h6901P}
\bibinfo{author}{\bibfnamefont{J.~R.} \bibnamefont{{Pritchard}}}
  \bibnamefont{and} \bibinfo{author}{\bibfnamefont{A.}~\bibnamefont{{Loeb}}},
  \bibinfo{journal}{Reports on Progress in Physics}
  \textbf{\bibinfo{volume}{75}}, \bibinfo{eid}{086901} (\bibinfo{year}{2012}),
  \eprint{1109.6012}.

\bibitem[{\citenamefont{{Oyama} et~al.}(2013)\citenamefont{{Oyama}, {Shimizu},
  and {Kohri}}}]{2013PhLB..718.1186O}
\bibinfo{author}{\bibfnamefont{Y.}~\bibnamefont{{Oyama}}},
  \bibinfo{author}{\bibfnamefont{A.}~\bibnamefont{{Shimizu}}},
  \bibnamefont{and} \bibinfo{author}{\bibfnamefont{K.}~\bibnamefont{{Kohri}}},
  \bibinfo{journal}{Physics Letters B} \textbf{\bibinfo{volume}{718}},
  \bibinfo{pages}{1186} (\bibinfo{year}{2013}), \eprint{1205.5223}.

\bibitem[{\citenamefont{{Kohri} et~al.}(2014)\citenamefont{{Kohri}, {Oyama},
  {Sekiguchi}, and {Takahashi}}}]{2014JCAP...09..014K}
\bibinfo{author}{\bibfnamefont{K.}~\bibnamefont{{Kohri}}},
  \bibinfo{author}{\bibfnamefont{Y.}~\bibnamefont{{Oyama}}},
  \bibinfo{author}{\bibfnamefont{T.}~\bibnamefont{{Sekiguchi}}},
  \bibnamefont{and}
  \bibinfo{author}{\bibfnamefont{T.}~\bibnamefont{{Takahashi}}},
  \bibinfo{journal}{\jcap} \textbf{\bibinfo{volume}{9}}, \bibinfo{eid}{014}
  (\bibinfo{year}{2014}), \eprint{1404.4847}.

\bibitem[{\citenamefont{{Shimabukuro} et~al.}(2014)\citenamefont{{Shimabukuro},
  {Ichiki}, {Inoue}, and {Yokoyama}}}]{2014PhRvD..90h3003S}
\bibinfo{author}{\bibfnamefont{H.}~\bibnamefont{{Shimabukuro}}},
  \bibinfo{author}{\bibfnamefont{K.}~\bibnamefont{{Ichiki}}},
  \bibinfo{author}{\bibfnamefont{S.}~\bibnamefont{{Inoue}}}, \bibnamefont{and}
  \bibinfo{author}{\bibfnamefont{S.}~\bibnamefont{{Yokoyama}}},
  \bibinfo{journal}{\prd} \textbf{\bibinfo{volume}{90}}, \bibinfo{eid}{083003}
  (\bibinfo{year}{2014}), \eprint{1403.1605}.

\bibitem[{MWA()}]{MWA}
\emph{\bibinfo{title}{\url{http://www.mwatelescope.org}}}.

\bibitem[{SKA()}]{SKA}
\emph{\bibinfo{title}{\url{http://www.skatelescope.org}}}.

\bibitem[{\citenamefont{{Tseliakhovich} and
  {Hirata}}(2010)}]{2010PhRvD..82h3520T}
\bibinfo{author}{\bibfnamefont{D.}~\bibnamefont{{Tseliakhovich}}}
  \bibnamefont{and} \bibinfo{author}{\bibfnamefont{C.}~\bibnamefont{{Hirata}}},
  \bibinfo{journal}{\prd} \textbf{\bibinfo{volume}{82}}, \bibinfo{eid}{083520}
  (\bibinfo{year}{2010}), \eprint{1005.2416}.

\bibitem[{\citenamefont{{Stacy} et~al.}(2011)\citenamefont{{Stacy}, {Bromm},
  and {Loeb}}}]{2011ApJ...730L...1S}
\bibinfo{author}{\bibfnamefont{A.}~\bibnamefont{{Stacy}}},
  \bibinfo{author}{\bibfnamefont{V.}~\bibnamefont{{Bromm}}}, \bibnamefont{and}
  \bibinfo{author}{\bibfnamefont{A.}~\bibnamefont{{Loeb}}},
  \bibinfo{journal}{\apjl} \textbf{\bibinfo{volume}{730}}, \bibinfo{eid}{L1}
  (\bibinfo{year}{2011}), \eprint{1011.4512}.

\bibitem[{\citenamefont{{Naoz} et~al.}(2012)\citenamefont{{Naoz}, {Yoshida},
  and {Gnedin}}}]{2012ApJ...747..128N}
\bibinfo{author}{\bibfnamefont{S.}~\bibnamefont{{Naoz}}},
  \bibinfo{author}{\bibfnamefont{N.}~\bibnamefont{{Yoshida}}},
  \bibnamefont{and} \bibinfo{author}{\bibfnamefont{N.~Y.}
  \bibnamefont{{Gnedin}}}, \bibinfo{journal}{\apj}
  \textbf{\bibinfo{volume}{747}}, \bibinfo{eid}{128} (\bibinfo{year}{2012}),
  \eprint{1108.5176}.

\bibitem[{\citenamefont{{Naoz} et~al.}(2013)\citenamefont{{Naoz}, {Yoshida},
  and {Gnedin}}}]{2013ApJ...763...27N}
\bibinfo{author}{\bibfnamefont{S.}~\bibnamefont{{Naoz}}},
  \bibinfo{author}{\bibfnamefont{N.}~\bibnamefont{{Yoshida}}},
  \bibnamefont{and} \bibinfo{author}{\bibfnamefont{N.~Y.}
  \bibnamefont{{Gnedin}}}, \bibinfo{journal}{\apj}
  \textbf{\bibinfo{volume}{763}}, \bibinfo{eid}{27} (\bibinfo{year}{2013}),
  \eprint{1207.5515}.

\bibitem[{\citenamefont{{Fialkov}}(2014)}]{2014IJMPD..2330017F}
\bibinfo{author}{\bibfnamefont{A.}~\bibnamefont{{Fialkov}}},
  \bibinfo{journal}{International Journal of Modern Physics D}
  \textbf{\bibinfo{volume}{23}}, \bibinfo{eid}{1430017} (\bibinfo{year}{2014}),
  \eprint{1407.2274}.

\bibitem[{\citenamefont{{Naoz} and {Barkana}}(2005)}]{2005MNRAS.362.1047N}
\bibinfo{author}{\bibfnamefont{S.}~\bibnamefont{{Naoz}}} \bibnamefont{and}
  \bibinfo{author}{\bibfnamefont{R.}~\bibnamefont{{Barkana}}},
  \bibinfo{journal}{\mnras} \textbf{\bibinfo{volume}{362}},
  \bibinfo{pages}{1047} (\bibinfo{year}{2005}), \eprint{astro-ph/0503196}.

\bibitem[{\citenamefont{{Tseliakhovich}
  et~al.}(2011)\citenamefont{{Tseliakhovich}, {Barkana}, and
  {Hirata}}}]{2011MNRAS.418..906T}
\bibinfo{author}{\bibfnamefont{D.}~\bibnamefont{{Tseliakhovich}}},
  \bibinfo{author}{\bibfnamefont{R.}~\bibnamefont{{Barkana}}},
  \bibnamefont{and} \bibinfo{author}{\bibfnamefont{C.~M.}
  \bibnamefont{{Hirata}}}, \bibinfo{journal}{\mnras}
  \textbf{\bibinfo{volume}{418}}, \bibinfo{pages}{906} (\bibinfo{year}{2011}),
  \eprint{1012.2574}.

\bibitem[{\citenamefont{{Naoz} et~al.}(2011)\citenamefont{{Naoz}, {Yoshida},
  and {Barkana}}}]{2011MNRAS.416..232N}
\bibinfo{author}{\bibfnamefont{S.}~\bibnamefont{{Naoz}}},
  \bibinfo{author}{\bibfnamefont{N.}~\bibnamefont{{Yoshida}}},
  \bibnamefont{and}
  \bibinfo{author}{\bibfnamefont{R.}~\bibnamefont{{Barkana}}},
  \bibinfo{journal}{\mnras} \textbf{\bibinfo{volume}{416}},
  \bibinfo{pages}{232} (\bibinfo{year}{2011}), \eprint{1009.0945}.

\bibitem[{\citenamefont{{Springel}}(2005)}]{2005MNRAS.364.1105S}
\bibinfo{author}{\bibfnamefont{V.}~\bibnamefont{{Springel}}},
  \bibinfo{journal}{\mnras} \textbf{\bibinfo{volume}{364}},
  \bibinfo{pages}{1105} (\bibinfo{year}{2005}), \eprint{astro-ph/0505010}.

\bibitem[{\citenamefont{{Worrakitpoonpon}}(2015)}]{2015MNRAS.446.1335W}
\bibinfo{author}{\bibfnamefont{T.}~\bibnamefont{{Worrakitpoonpon}}},
  \bibinfo{journal}{\mnras} \textbf{\bibinfo{volume}{446}},
  \bibinfo{pages}{1335} (\bibinfo{year}{2015}), \eprint{1410.4272}.

\bibitem[{\citenamefont{{Planck Collaboration}
  et~al.}(2014)\citenamefont{{Planck Collaboration}, {Ade}, {Aghanim},
  {Armitage-Caplan}, {Arnaud}, {Ashdown}, {Atrio-Barandela}, {Aumont},
  {Baccigalupi}, {Banday} et~al.}}]{2014A&A...571A..16P}
\bibinfo{author}{\bibnamefont{{Planck Collaboration}}},
  \bibinfo{author}{\bibfnamefont{P.~A.~R.} \bibnamefont{{Ade}}},
  \bibinfo{author}{\bibfnamefont{N.}~\bibnamefont{{Aghanim}}},
  \bibinfo{author}{\bibfnamefont{C.}~\bibnamefont{{Armitage-Caplan}}},
  \bibinfo{author}{\bibfnamefont{M.}~\bibnamefont{{Arnaud}}},
  \bibinfo{author}{\bibfnamefont{M.}~\bibnamefont{{Ashdown}}},
  \bibinfo{author}{\bibfnamefont{F.}~\bibnamefont{{Atrio-Barandela}}},
  \bibinfo{author}{\bibfnamefont{J.}~\bibnamefont{{Aumont}}},
  \bibinfo{author}{\bibfnamefont{C.}~\bibnamefont{{Baccigalupi}}},
  \bibinfo{author}{\bibfnamefont{A.~J.} \bibnamefont{{Banday}}},
  \bibnamefont{et~al.}, \bibinfo{journal}{\aap} \textbf{\bibinfo{volume}{571}},
  \bibinfo{eid}{A16} (\bibinfo{year}{2014}), \eprint{1303.5076}.

\bibitem[{\citenamefont{{Lewis} et~al.}(2000)\citenamefont{{Lewis},
  {Challinor}, and {Lasenby}}}]{2000ApJ...538..473L}
\bibinfo{author}{\bibfnamefont{A.}~\bibnamefont{{Lewis}}},
  \bibinfo{author}{\bibfnamefont{A.}~\bibnamefont{{Challinor}}},
  \bibnamefont{and}
  \bibinfo{author}{\bibfnamefont{A.}~\bibnamefont{{Lasenby}}},
  \bibinfo{journal}{\apj} \textbf{\bibinfo{volume}{538}}, \bibinfo{pages}{473}
  (\bibinfo{year}{2000}), \eprint{astro-ph/9911177}.

\bibitem[{\citenamefont{{Naoz} and {Narayan}}(2014)}]{2014ApJ...791L...8N}
\bibinfo{author}{\bibfnamefont{S.}~\bibnamefont{{Naoz}}} \bibnamefont{and}
  \bibinfo{author}{\bibfnamefont{R.}~\bibnamefont{{Narayan}}},
  \bibinfo{journal}{\apjl} \textbf{\bibinfo{volume}{791}}, \bibinfo{eid}{L8}
  (\bibinfo{year}{2014}), \eprint{1407.3795}.

\bibitem[{\citenamefont{{Greif} et~al.}(2011)\citenamefont{{Greif}, {White},
  {Klessen}, and {Springel}}}]{2011ApJ...736..147G}
\bibinfo{author}{\bibfnamefont{T.~H.} \bibnamefont{{Greif}}},
  \bibinfo{author}{\bibfnamefont{S.~D.~M.} \bibnamefont{{White}}},
  \bibinfo{author}{\bibfnamefont{R.~S.} \bibnamefont{{Klessen}}},
  \bibnamefont{and}
  \bibinfo{author}{\bibfnamefont{V.}~\bibnamefont{{Springel}}},
  \bibinfo{journal}{\apj} \textbf{\bibinfo{volume}{736}}, \bibinfo{eid}{147}
  (\bibinfo{year}{2011}), \eprint{1101.5493}.

\bibitem[{\citenamefont{{Maio} et~al.}(2011)\citenamefont{{Maio}, {Koopmans},
  and {Ciardi}}}]{2011MNRAS.412L..40M}
\bibinfo{author}{\bibfnamefont{U.}~\bibnamefont{{Maio}}},
  \bibinfo{author}{\bibfnamefont{L.~V.~E.} \bibnamefont{{Koopmans}}},
  \bibnamefont{and} \bibinfo{author}{\bibfnamefont{B.}~\bibnamefont{{Ciardi}}},
  \bibinfo{journal}{\mnras} \textbf{\bibinfo{volume}{412}},
  \bibinfo{pages}{L40} (\bibinfo{year}{2011}), \eprint{1011.4006}.

\bibitem[{\citenamefont{{Fialkov} et~al.}(2012)\citenamefont{{Fialkov},
  {Barkana}, {Tseliakhovich}, and {Hirata}}}]{2012MNRAS.424.1335F}
\bibinfo{author}{\bibfnamefont{A.}~\bibnamefont{{Fialkov}}},
  \bibinfo{author}{\bibfnamefont{R.}~\bibnamefont{{Barkana}}},
  \bibinfo{author}{\bibfnamefont{D.}~\bibnamefont{{Tseliakhovich}}},
  \bibnamefont{and} \bibinfo{author}{\bibfnamefont{C.~M.}
  \bibnamefont{{Hirata}}}, \bibinfo{journal}{\mnras}
  \textbf{\bibinfo{volume}{424}}, \bibinfo{pages}{1335} (\bibinfo{year}{2012}),
  \eprint{1110.2111}.

\bibitem[{\citenamefont{{Ferraro} et~al.}(2012)\citenamefont{{Ferraro},
  {Smith}, and {Dvorkin}}}]{2012PhRvD..85d3523F}
\bibinfo{author}{\bibfnamefont{S.}~\bibnamefont{{Ferraro}}},
  \bibinfo{author}{\bibfnamefont{K.~M.} \bibnamefont{{Smith}}},
  \bibnamefont{and}
  \bibinfo{author}{\bibfnamefont{C.}~\bibnamefont{{Dvorkin}}},
  \bibinfo{journal}{\prd} \textbf{\bibinfo{volume}{85}}, \bibinfo{eid}{043523}
  (\bibinfo{year}{2012}), \eprint{1110.2182}.

\bibitem[{\citenamefont{{Bittner} and {Loeb}}(2011)}]{2011arXiv1110.4659B}
\bibinfo{author}{\bibfnamefont{J.~M.} \bibnamefont{{Bittner}}}
  \bibnamefont{and} \bibinfo{author}{\bibfnamefont{A.}~\bibnamefont{{Loeb}}},
  \bibinfo{journal}{ArXiv e-prints}  (\bibinfo{year}{2011}),
  \eprint{1110.4659}.

\bibitem[{\citenamefont{{McQuinn} and {O'Leary}}(2012)}]{2012ApJ...760....3M}
\bibinfo{author}{\bibfnamefont{M.}~\bibnamefont{{McQuinn}}} \bibnamefont{and}
  \bibinfo{author}{\bibfnamefont{R.~M.} \bibnamefont{{O'Leary}}},
  \bibinfo{journal}{\apj} \textbf{\bibinfo{volume}{760}}, \bibinfo{eid}{3}
  (\bibinfo{year}{2012}), \eprint{1204.1345}.

\bibitem[{\citenamefont{{Tashiro} et~al.}(2008)\citenamefont{{Tashiro},
  {Aghanim}, {Langer}, {Douspis}, and {Zaroubi}}}]{2008MNRAS.389..469T}
\bibinfo{author}{\bibfnamefont{H.}~\bibnamefont{{Tashiro}}},
  \bibinfo{author}{\bibfnamefont{N.}~\bibnamefont{{Aghanim}}},
  \bibinfo{author}{\bibfnamefont{M.}~\bibnamefont{{Langer}}},
  \bibinfo{author}{\bibfnamefont{M.}~\bibnamefont{{Douspis}}},
  \bibnamefont{and}
  \bibinfo{author}{\bibfnamefont{S.}~\bibnamefont{{Zaroubi}}},
  \bibinfo{journal}{\mnras} \textbf{\bibinfo{volume}{389}},
  \bibinfo{pages}{469} (\bibinfo{year}{2008}), \eprint{0802.3893}.

\bibitem[{\citenamefont{{Yoo} et~al.}(2011)\citenamefont{{Yoo}, {Dalal}, and
  {Seljak}}}]{2011JCAP...07..018Y}
\bibinfo{author}{\bibfnamefont{J.}~\bibnamefont{{Yoo}}},
  \bibinfo{author}{\bibfnamefont{N.}~\bibnamefont{{Dalal}}}, \bibnamefont{and}
  \bibinfo{author}{\bibfnamefont{U.}~\bibnamefont{{Seljak}}},
  \bibinfo{journal}{\jcap} \textbf{\bibinfo{volume}{7}}, \bibinfo{eid}{018}
  (\bibinfo{year}{2011}), \eprint{1105.3732}.

\bibitem[{\citenamefont{{Slepian} and
  {Eisenstein}}(2015)}]{2015MNRAS.448....9S}
\bibinfo{author}{\bibfnamefont{Z.}~\bibnamefont{{Slepian}}} \bibnamefont{and}
  \bibinfo{author}{\bibfnamefont{D.~J.} \bibnamefont{{Eisenstein}}},
  \bibinfo{journal}{\mnras} \textbf{\bibinfo{volume}{448}}, \bibinfo{pages}{9}
  (\bibinfo{year}{2015}), \eprint{1411.4052}.

\bibitem[{\citenamefont{{Yoo} and {Seljak}}(2013)}]{2013PhRvD..88j3520Y}
\bibinfo{author}{\bibfnamefont{J.}~\bibnamefont{{Yoo}}} \bibnamefont{and}
  \bibinfo{author}{\bibfnamefont{U.}~\bibnamefont{{Seljak}}},
  \bibinfo{journal}{\prd} \textbf{\bibinfo{volume}{88}}, \bibinfo{eid}{103520}
  (\bibinfo{year}{2013}), \eprint{1308.1401}.

\bibitem[{\citenamefont{{Beutler} et~al.}(2015)\citenamefont{{Beutler},
  {Blake}, {Koda}, {Marin}, {Seo}, {Cuesta}, and
  {Schneider}}}]{2015arXiv150603900B}
\bibinfo{author}{\bibfnamefont{F.}~\bibnamefont{{Beutler}}},
  \bibinfo{author}{\bibfnamefont{C.}~\bibnamefont{{Blake}}},
  \bibinfo{author}{\bibfnamefont{J.}~\bibnamefont{{Koda}}},
  \bibinfo{author}{\bibfnamefont{F.}~\bibnamefont{{Marin}}},
  \bibinfo{author}{\bibfnamefont{H.-J.} \bibnamefont{{Seo}}},
  \bibinfo{author}{\bibfnamefont{A.~J.} \bibnamefont{{Cuesta}}},
  \bibnamefont{and} \bibinfo{author}{\bibfnamefont{D.~P.}
  \bibnamefont{{Schneider}}}, \bibinfo{journal}{ArXiv e-prints}
  (\bibinfo{year}{2015}), \eprint{1506.03900}.

\end{thebibliography}

\end{document}